\documentclass[iop]{emulateapj}

 \usepackage{tipa}
 \usepackage{amsmath,bm}
 
\usepackage{amssymb}
 \usepackage{hyperref}
\usepackage{natbib}
\slugcomment{Accepted for Publication in ApJL}

\shorttitle{`Oumuamua}
\shortauthors{Seligman, Laughlin $\&$ Batygin}

\begin{document}

\title{On the Anomalous Acceleration of 1I/2017 U1 `Oumuamua}

\author{Darryl Seligman and Gregory Laughlin}
\affil{Dept. of Astronomy, Yale University,
    New Haven, CT 06517}
\email{darryl.seligman@yale.edu}
    
\and

\author{Konstantin Batygin}
\affil{Division of Geological and Planetary Sciences, Caltech, Pasadena, CA 91125}

\begin{abstract}
We show that the $P\sim8\,{\rm h}$ photometric period and the astrometrically measured $A_{\rm ng}\sim2.5\times10^{-4}\,{\rm cm\,s^{-2}}$ non-gravitational acceleration (at $r\sim1.4\,{\rm AU}$) of the interstellar object 1I/2017 (`Oumuamua) can be explained by a nozzle-like venting of volatiles whose activity migrated to track the sub-solar location on the object's surface. Adopting the assumption that `Oumuamua was an elongated $a\times b \times c$ ellipsoid, this model produces a pendulum-like rotation of the body and implies a long semi-axis $a\sim 5A_{\rm ng}P^2/4\pi^2 \sim 260\,{\rm m}$. This scale agrees with the independent estimates of `Oumuamua's size that stem from its measured brightness, assuming an albedo of $p\sim0.1$, appropriate to ices that have undergone long-duration exposure to the interstellar cosmic ray flux. Using ray-tracing, we generate light curves for ellipsoidal bodies that are subject to both physically consistent sub-solar torques and to the time-varying geometry of the Sun-Earth-`Oumuamua configuration. Our synthetic light curves display variations from chaotic tumbling and changing cross-sectional illumination that are consistent with the observations, while avoiding significant secular changes in the photometric periodicity. If our model is correct, `Oumuamua experienced mass loss that wasted $\sim 10\%$ of its total mass during the $\sim 100\,{\rm d}$ span of its encounter with the inner Solar System  and had an icy composition with a very low $[{\rm C}/{\rm O}]\lesssim 0.003$. Our interpretation of `Oumuamua's behavior  is consistent with the hypothesis that it was ejected from either the outer regions of a planetesimal disk after an encounter with an embedded $M_{\rm p} \sim M_{\rm Nep}$ planet  or from an exo-Oort cloud.
\end{abstract}


\section{Introduction}
`Oumuamua  was
the first macroscopic object of clear interstellar origin to be seen within the Solar System. Its appearance was unexpected, and its behavior defied expectations. Studies, including those by \citet{Moro2009} and \citet{Cook2016} generated an assessment (resting on small-body size distribution estimates and exoplanet occurrence rates) that interstellar objects would be found only when the Large Synoptic Survey Telescope is operational.

The observational facts are readily summarized. `Oumuamua arrived from the direction of the galactic apex on a hyperbolic trajectory, with $v_{\inf}\sim26\,{\rm km\,s^{-1}}$, a value similar to the local velocity dispersion of Population I stars \citep{mam17}. After experiencing a close approach, $q\sim0.25\,{\rm AU}$ to the Sun on 9 Sep, 2017, `Oumuamua was discovered post-periastron on 19 Oct, 2017 by Pan-STARRS \citep{williams2017}. A variety of observational campaigns were quickly organized on telescopes worldwide, generating a high-quality composite light curve comprising 818 observations and spanning 29.3 days, as summarized by \citet{bel18}. Frequency analysis of the light curve shows a power maximum at $P\sim4.3\,{\rm hr}$, which was interpreted to be half the spin period of a rotating body. `Oumuamua's light curve exhibited irregular, $f=F_{\rm max}/F_{\rm min}\sim15$, flux variations that were explained by positing an elongated shape experiencing complex, non-principle axis rotation \citep{mee17, fra18, dra18}.

\citet{mic18} analyzed all extant photometry for `Oumuamua (including multiple HST observations taken through the end of 2017) and determined that its outbound trajectory was strongly inconsistent with motion subject only to solar gravity. \citet{mic18} determined that a radially outward acceleration component of functional form ${\bf \alpha}=4.92\times10^{-4}(r/1 {\rm AU})^{-2}\,\hat{\bf r}\,{\rm cm\,s^{-2}}$ superimposed on the Keplerian acceleration permits a much better fit to the observed trajectory. The required magnitude of non-gravitational acceleration, $A_{\rm ng}\sim 2.5\times10^{-4}\,{\rm cm\,s^{-2}}$ at $r\sim 1.4\,{\rm AU}$ where `Oumuamua was observed at highest signal-to-noise, is of order $10^{-3}$ of the solar gravitational acceleration.

\citet{mic18} concluded that directed out-gassing from the surface \citep[e.g.][]{mar73} is the most viable explanatory mechanism, with the model requiring a mass flux of $\dot{m}\sim10^4\,{\rm g\,s^{-1}}$ jetting in the solar direction at $v\sim3\times10^4\,{\rm cm\,s^{-1}}$. 
An outflow of this order of magnitude is not unusual for comets \citep[see, e.g.][]{Lee2015}, but is curious in light of `Oumuamua's small inferred mass ($M\sim10^{9}{\rm kg}$),  absence of an observed coma entraining $\mu$m-sized dust \citep{mee17,Jewitt2017}, and the  non-detection of carbon-containing outgassed species including CN, CO and ${\rm CO}_{2}$ \citep{Ye2017, Trilling2018}.\citet{Rafikov2018}, moreover, showed that a traditional cometary jet interpretation of the non-gravitational acceleration, where the reactive torques from the jet,  averaged over the trajectory, increase the angular momentum of the body by an amount set by a  dimensionless lever arm parameter, is problematic. Torques associated with a jet similar to those seen on Solar System comets would have observably spun up the body during the period over which it was monitored. 

We propose that -- even in light of the dust, gas-composition and spin-up issues -- a volatile-rich gas-venting structure for `Oumuamua provides the simplest explanation for its odd trajectory. Alternate models that invoke explosive break-up provide inferior fits to the astrometry \citep{mic18}, whereas explanations that invoke radiation pressure require an unusual physical  geometry  \citep{Bialy2018},  or internal structure \citep{moro2019b} for the body.

Here, we argue that the venting of heat-mobilized near-surface volatiles can simultaneously explain both `Oumuamua's light curve and its acceleration. Treating `Oumuamua as a monolithic tri-axial ellipsoid, we model its solid body dynamics under the assumption that a jet directed normally to the surface tracks the spot of maximum insolation. We first show that a scenario of this type is broadly consistent with the observations, and we then  briefly discuss the ramifications.

\section{Dynamical Model}
\begin{figure}
\includegraphics[trim=0.5cm 0cm 0.5cm 0cm,scale=.25]{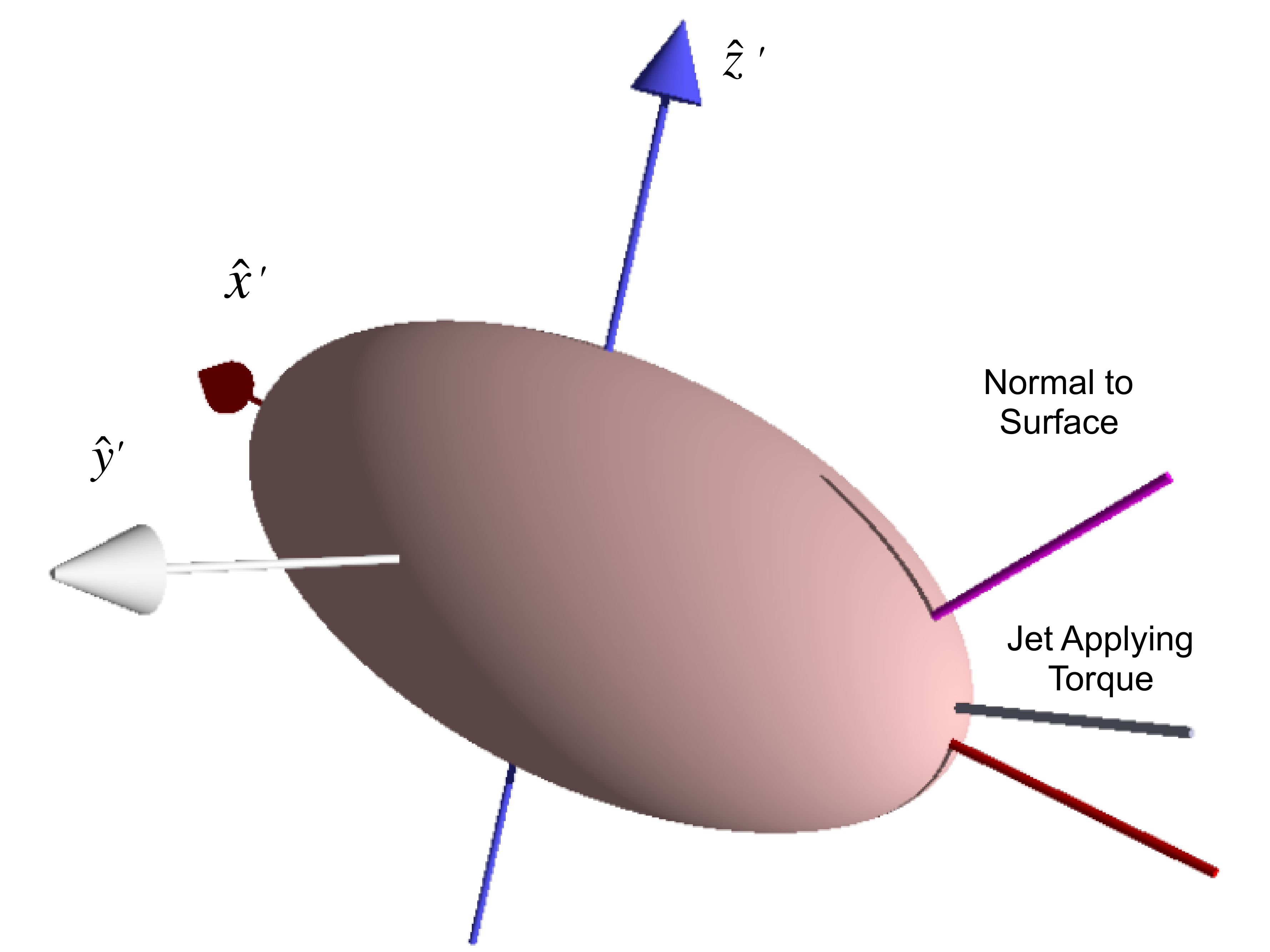}
\caption{Geometry of our model as rendered with ray tracing. The line labeled ``Jet Applying Torque'' shows the ${\hat x}$ direction and thus casts no shadow. The vectors ${\hat x}^\prime$, ${\hat y}^\prime$, and ${\hat z}^\prime$ point, respectively, along the principal axes of an $a \times b \times c$ ellipsoid. The shadow cast by an example vector normal to the illuminated ellipsoidal surface is also shown. For clarity of illustration, the figure adopts $a=2, b=1, c=1$. \label{schematic}}
\end{figure}

Consider a jet that migrates to track the substellar point on the surface of an illuminated ellipsoid. A model of this type has been successfully applied to comet 67P/Churyumov-Gerasimenko \citep[see, e.g.][]{kra19}. The jet vents in the direction, $\hat{n}$, normal to the surface, thereby exerting the non-gravitational force that \citet{mic18} find provides the best model fit to `Oumuamua's astrometry
\begin{equation}
F(t)\hat{\bf n}=4.92\times10^{-4}\,M\,r(t)^{-2}\hat{\bf r}\,{\rm cm\,s^{-2}}\,,
\end{equation}
where $M$ is the ellipsoid's mass, and $r(t)$ is the radial distance from the Sun in AU. We wish to calculate the net effect of the jet producing this acceleration on the evolving rotational state of the body. 

We shift to a non-inertial frame that co-moves with the ellipsoid, whose center of mass is taken as the origin. We define $\hat x$ as the direction of the radial vector connecting the Sun to the ellipsoid's center of mass. The sub-stellar point of maximal irradiation occurs where ${\hat x} \times {\hat n}=0$. Figure \ref{schematic} provides a schematic.

We first consider a restricted situation in which the shortest axis of the ellipsoid is aligned with $\hat z$. The jet-induced torque then acts only in the $x$-$y$ plane and the ellipsoid has a minimum projected semi-minor/major axis ratio, $\epsilon=b/a$. The angle $\theta\in(0,\pi)$ represents a rotation about the $z$-axis. Assuming the body starts with zero angular momentum, the idealized jet never induces a full rotation, e.g. $\theta\not\ge\pi$. When $\theta=0$, the semi-major/minor axis lies along the $x$-/$y$-axis. To locate the sub-stellar point, we construct the ellipsoid in the co-moving non-rotating  frame,  
\begin{equation}
    f(x,y)=\frac{x^2}{a^2}+\frac{y^2}{b^2}-1=0 \, .\label{ellipsoidequation2d}
\end{equation}
The unit normal to this ellipsoid is given by
\begin{equation}
   \frac{{\bf{\nabla}}f}{|{\bf{\nabla}}f|}=
    \left[\frac{x^2}{a^4}+\frac{y^2}{b^4}\right]^{-1/2}
    \left[\frac{x}{a^2}\hat{\bf i}+\frac{y}{b^2}\hat{\bf j}\right]\, .\label{gradientEq2d}
\end{equation}
The angle $\theta$ defines a rotation matrix, which we use to rotate the vector to the Sun by $\theta$, so that in the rotated frame, the Sun shines along the direction defined by $\cos(\theta)\hat{\bf i}+\sin(\theta)\hat{\bf j}$. 

The system is thus described by a simple Hamiltonian with one degree of freedom,

\begin{equation}
\mathcal{H}=\frac{\Theta^2}{2}-\omega_0^2\frac{\sqrt{1+\epsilon^2-(\epsilon^2-1)\cos(2\theta)}}{\sqrt2(\epsilon^2-1)}
    \,,\label{Hamiltonian}
\end{equation}
where $\omega_0$ is defined in Equation \ref{frequency_omega0} below. With the relevant moment of inertia, $I=Ma^2(1+\epsilon^2)/5$, the equation of motion is

\begin{equation}
    \frac{d^2\theta}{dt^2} = \omega_0^2\frac{\cos(\theta)\sin{\theta}}{\sqrt{\cos^2(\theta) +\sin^2(\theta) \epsilon^2}} \,,\label{EOM}
\end{equation}
where
\begin{equation}
     \omega_0^2=\frac{5A_{\rm ng}}{a}\frac{(1-\epsilon^2)}{(1+\epsilon^2)}\,.\label{frequency_omega0}
\end{equation}

For small $\epsilon$, the Hamiltonian reduces to
\begin{equation}
\mathcal{H}\approx\frac{\Theta^2}{2}-\omega_0^2|\cos(\theta)|+\mathcal{O}(\epsilon^2)\,.\label{Hamiltonian_reduced}
\end{equation}
To second order in $\epsilon$, the Hamiltonian depends only on $a$, and hence
\begin{equation}
\omega_0^2\sim\frac{5A_{\rm ng}}{a}\,.\label{omega_0}
\end{equation}

\begin{figure}
\includegraphics[trim=0.0cm 0cm 3.0cm 0.0cm,scale=.52]{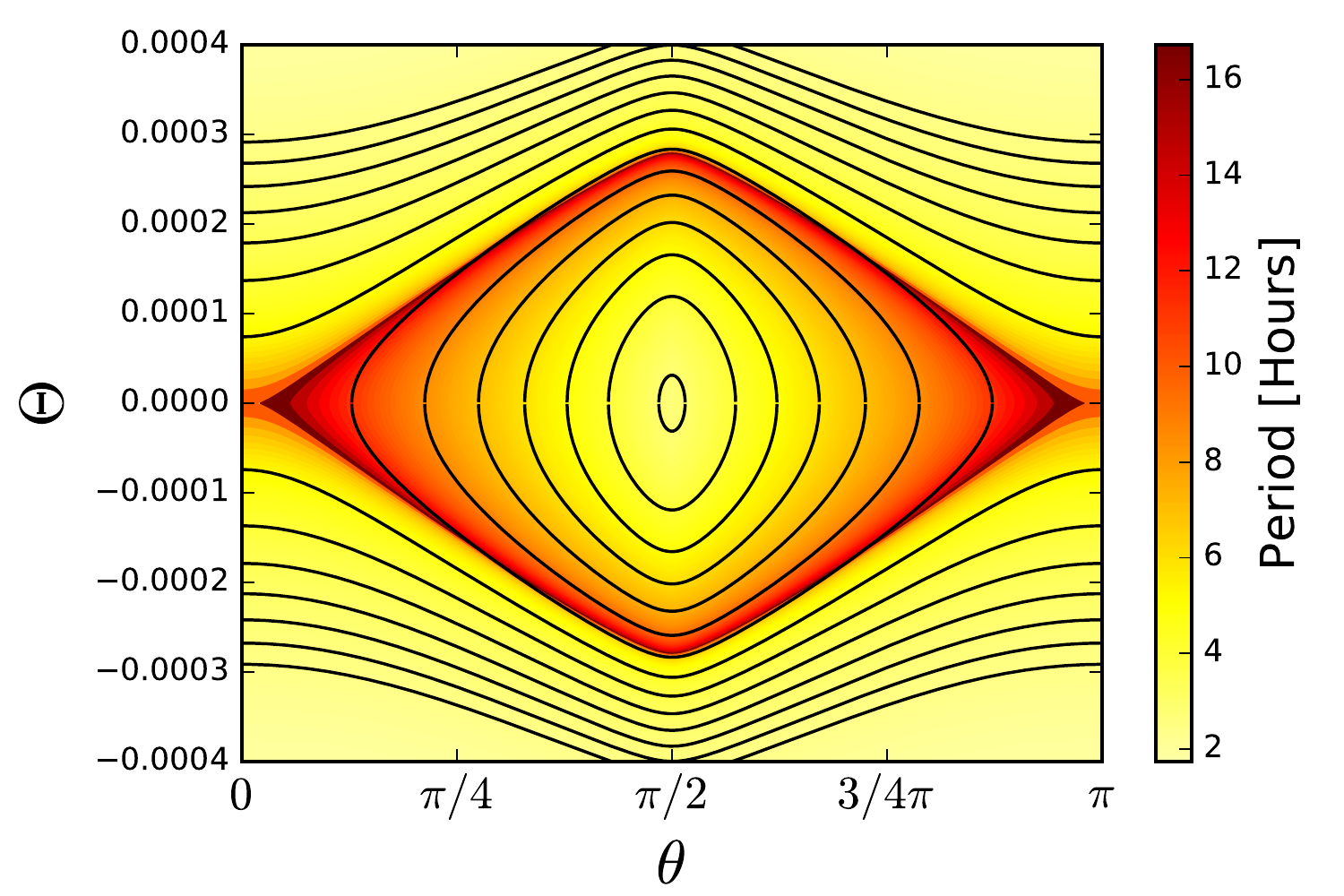}
\caption{Phase space diagram of the motion described  by Equation \ref{EOM} for an ellipsoid with an aspect ratio, $a$:$b$, of $9$:$1$. The colorscale shows the period of libration or circulation. For substantial range of initial venting angles, the period of oscillations is close to the observed $P\sim8{\rm \,hr}$. \label{phasespace}}
\end{figure}

Figure \ref{phasespace} shows the level curves of the Hamiltionian, given by Equation \ref{Hamiltonian}, for an ellipsoid with an aspect ratio,  $a$:$b$, of $9$:$1$. For a substantial range of initial jet angles, the period of oscillations is close to $P\sim8{\rm \,hr}$.

The period of oscillation depends strongly on the length, $a$, of the long axis, and on the magnitude of the  acceleration, $A_{\rm ng}$, but only weakly on the aspect ratio, $\epsilon$. The observed light curve therefore implies a length scale $a\sim 5A_{\rm ng}P^2/4\pi^2 \sim 260\,{\rm m}$. This scale concords with the independent estimates of `Oumuamua's size that stem from its measured brightness \citep[e.g.][]{mee17, Jewitt2017} if we assume an albedo, $p\sim0.1$, appropriate to surface ices that have undergone long-duration exposure to the interstellar cosmic ray flux \citep{Moore1983}. This albedo and this size are also consistent with the non-detection of `Oumuamua in the infrared using the Spitzer Space Telescope \citep{Trilling2018} which implies an effective radius, $49\,{\rm m}<R_{\rm eff} =({\sigma/\pi})^{1/2}<220\,{\rm m}$, depending on the nature of the surface (and where $\sigma$ is the cross-sectional area).

\subsection{3-Dimensional Dynamics}

In three dimensions, the motion is more complex.  The rotational state of the body is described by a rotation matrix, ${\bf R}$, and an angular momentum vector, ${\bf L}=(L_x,L_y,L_z)^T$. At time, $t$, the orientation of the Sun (${\hat x}$) with respect to the principal axes (${\hat x}^{\prime}, {\hat y}^{\prime}, {\hat z}^{\prime}$) of `Oumuamua is defined by ${\bf R}$. 

The equation for the ellipsoid in its body frame is
\begin{equation}
    f(x',y',z')=\frac{x'^2}{a^2}+\frac{y'^2}{b^2}+\frac{z'^2}{c^2}-1=0 \, ,
\end{equation}
We rotate the unit vector pointing to the Sun, so that the direction of illumination is defined by 
\begin{equation}
    R^{-1}_{xx}\hat{\bf i}+R^{-1}_{yx}\hat{\bf j}+R^{-1}_{zx}\hat{\bf k}={\bf R^{-1}}\begin{pmatrix}1\\0\\0\end{pmatrix}\,\label{SunUnitVector}.
\end{equation}
The unit normal to the surface is
\begin{equation}
   \frac{{\bf{\nabla}}f}{|{\bf{\nabla}}f|}=
    \left[\frac{x'^2}{a^4}+\frac{y'^2}{b^4}+\frac{z'^2}{c^4}\right]^{-1/2}
    \left[\frac{x'}{a^2}\hat{\bf i}+\frac{y'}{b^2}\hat{\bf j}+\frac{z'}{c^2}\hat{\bf k}\right]\, .\label{gradientEq}
\end{equation}
Equating expressions \ref{SunUnitVector} and \ref{gradientEq} permits expression of the substellar point, ($x'_{ss}$, $y'_{ss}$, $z'_{ss}$) in terms of $R^{-1}_{xx}$, $R^{-1}_{yx}$ and $Rv^{-1}_{zx}$. Upon defining the normalization factor, 
\begin{equation}
f = \sqrt{(R^{-1}_{xx}a)^2+(R^{-1}_{yx}b)^2+(R^{-1}_{zx}c)^2}\,, 
\end{equation}
the two solutions to the quadratic system of equations are
\begin{equation}
\begin{gathered}
    x'_{ss} =\pm R^{-1}_{xx}a^2 /f \\ 
    y'_{ss} =\pm R^{-1}_{yx}b^2 /f \\
    z'_{ss} =\pm R^{-1}_{zx}c^2 /f\,.
\end{gathered}
\end{equation}

This unit vector is normal to two unique points on the surface of the ellipsoid; the sub-stellar point and the antipodal point at the opposing surface. We thus require  
\begin{equation}
    R_{xx}^{-1}x'/a^2+ R_{yx}^{-1}y'/b^2+ R_{zx}^{-1}z'/c^2<0 \, ,
\end{equation}
giving the sub-stellar point for an ellipsoid centered in the co-moving non-rotating frame, and illuminated from the direction given by ${\bf R}^{-1}$. To find the actual sub-stellar point $P=(\xi,\eta,\zeta)$, we rotate the point on the ellipsoid using ${\bf R}$,
\begin{equation}
    (\xi,\eta,\zeta)^T={\bf R }(x'_{ss},y'_{ss},z'_{ss})^T \,.
\end{equation}

\begin{figure*}
\includegraphics[trim=2cm 0cm 0cm 0cm,scale=.54]{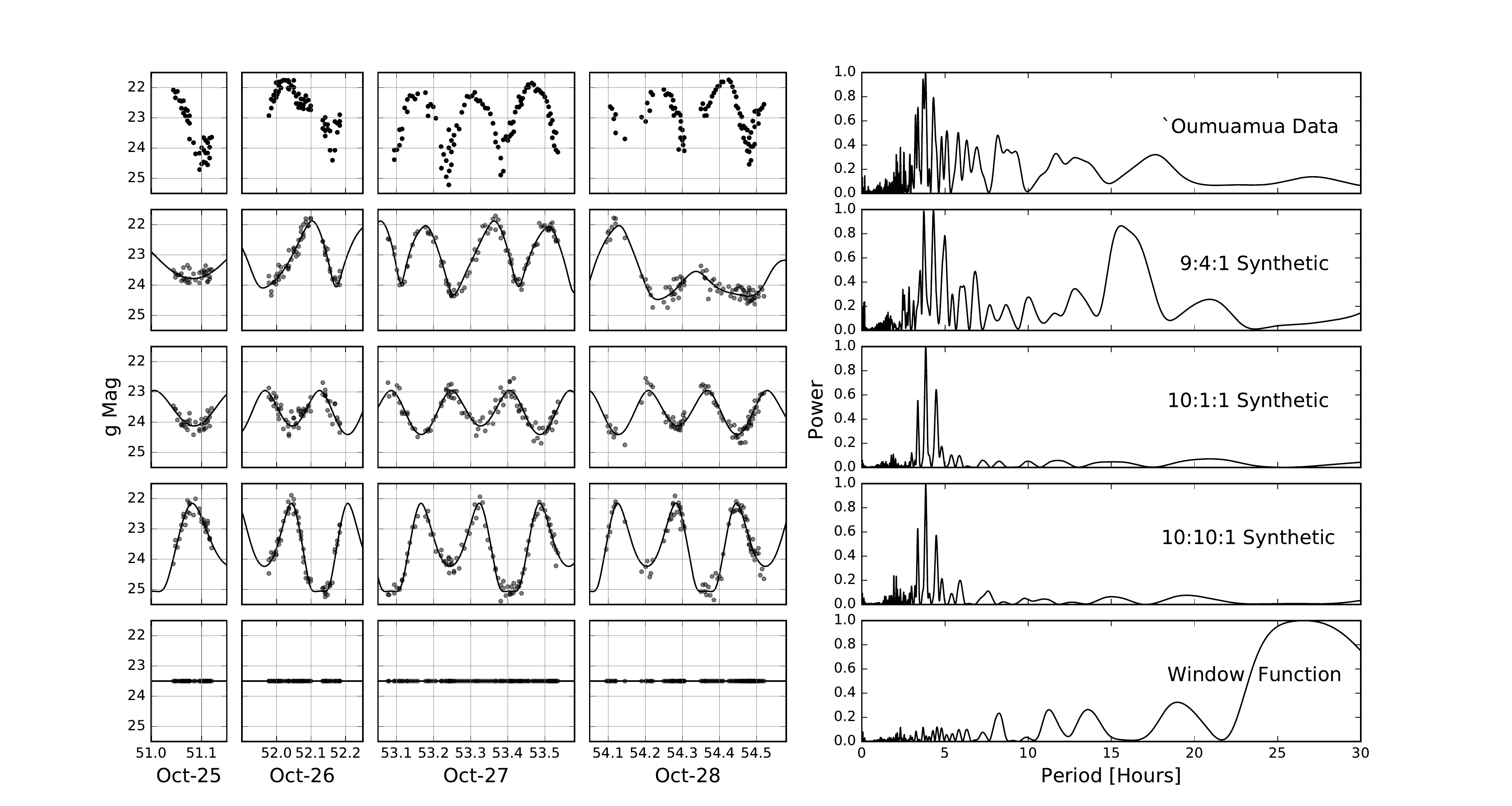}
\caption{Real and synthetic observations of `Oumuamua from October 25th-28th 2017 (\textit{left}) and their corresponding periodograms (\textit{right}). The rows show the real observations presented in \cite{bel18} (\textit{upper}), synthetic observations for the  9:4:1 (\textit{upper middle}) , 10:1:1 (\textit{middle}), and 10:10:1 (\textit{lower middle}) models, and a flat photometry (\textit{lower}). The solid lines show the underlying light curve for each model, and the transparent points show  synthetic observations sampled at the same epochs that `Oumuamua was observed, and perturbed with magnitude-dependant Gaussian noise inferred from the time series photometry. These synthetic measurements were used to compute the corresponding periodograms. 
\label{FullLightCurve}}
\end{figure*}

Given the sub-stellar point, we can compute the applied torque, ${\bm \tau}=(\xi,\eta,\zeta) \times {\bf F}$. The torque vector is thus
\begin{equation}
    {\bm \tau}(t)= MA_{\rm ng}( \zeta\hat{\bf j}-  \eta\hat{\bf k}) \, .
\end{equation}

With the ellipsoid treated as a rigid body \citep{gol50}, we integrate the coupled system of ordinary differential equations,
\begin{equation}
    \frac{d}{dt}{\bf L}(t)={\bm \tau}(t)\, ,
\end{equation}

and 
\begin{equation}
    \frac{d}{dt}{\bf R}_i(t)=[{\bf I}(t)^{-1}\cdot {\bf L}(t)]\times {\bf R}_i(t)\, .
\end{equation}
Here ${\bf R}_i(t)$ corresponds to the three column vectors of the ${\bf R}$, and ${\bf I}(t)$ is the time-dependent moment of inertia tensor, ${\bf I}(t)={\bf R}(t){\bf I}_{E} {\bf R}(t)^{\bf T}$, with ${\bf I}_{E}$, the tensor in the body frame of the ellipsoid, given by

\begin{equation}
    {\bf I}_{E}=
\frac{1}{5}M\begin{bmatrix} 
b^2+c^2 & 0 & 0 \\ 
0 & a^2 + c^2 & 0\\ 
0 & 0 & a^2+b^2
\end{bmatrix}\, .
\end{equation}

Given a choice for $a$:$b$:$c$, we integrate these eleven equations of motion\footnote{Within this construction, the angular momentum, $L_x$, in the $\hat{x}$ direction is constant. Therefore, the eleven equations of motion are for each component of the rotation matrix, and the angular momenta in the $\hat{y}$ and $\hat{z}$ directions, $L_y$ and $L_z$}, over the time span of the observed photometry, starting from a random choice for the initial rotation matrix and zero initial angular momentum. We include the radial dependence of the acceleration, but this makes a negligible difference over the time span considered. We verified that simulations that start at much earlier points in `Oumuamua's trajectory produced the same dynamics. Additionally, we validated our routine by checking that it recovers the level curves of the Hamiltonian describing the idealized 2-dimensional symmetry, as shown in Figure \ref{phasespace}.

We explored the effects of less idealized jet models, and found that they produced little difference to the rotational dynamics. We found that both (1) stochastic forcing, in which the applied force experienced random variations in magnitude of functional form,
\begin{equation}
    A_{\rm ng}(t+\delta t)=A_{\rm ng}(t) e^{-\delta t/\tau}+\Xi\sqrt{1-e^{-2\delta t/\tau}}
    \, ,\label{stocastic_forcing}
\end{equation}
where $\tau$ is the auto-correlation timescale and $\Xi$ is a random variable with normal distribution and a variance of unity \citep{Rein2010}, and (2) a time delay (subsolar lag of magnitudes ranging from $\tau_{lag}\sim P/6$ to $\tau_{lag}\sim P/2$) on the point where the force was applied, did not fundamentally change the resulting dynamics.



\subsection{Light Curve Rendering}
For a given initial condition, we produce a light curve using open source ray-tracing software\footnote{\url{http://www.povray.org/}}. We construct a scene with an ellipsoid of a given axis ratio, and place the camera (observer at Earth) and light source (Sun) in their correct positions. We calculate the illumination based on the diffuse reflectivity of the object, and we sum the brightness in the returned image to produce an unresolved flux. The flux is updated to generate a synthetic light curve as the body changes its orientation and as Earth and the body move through their known trajectories. 

In the left column of Figure \ref{FullLightCurve}, we show the real photometry and several sets of synthetic observations of `Oumumaua from October 25th-28th 2017, when the returned data quality was at its highest. The sample model geometries have dimensions $9$:$4$:$1$, $10$:$1$:$1$ and $10$:$10$:$1$. The third configuration invokes the SAM rotation state proposed by \cite{bel18}.
`Oumuamua's photometry was digitized from \cite{bel18} using \textit{Automeris} \citep{ank17}. 
The $\mathcal{O}(\epsilon^2)$ dependence implied for $\omega_{0}^{2}$ by Equation \ref{frequency_omega0} indicates that the observed values for $P$ and $A_{\rm ng}$ exert little constraint on the allowed range of aspect ratios. Constraints on $\epsilon$ arise primarily from the light curve variations. The aspect ratios that we have chosen are illustrative, and not the result of model optimization. Other values, such as the $6$:$1$ suggested by \citet{Jewitt2017} and \citet{Mcneill2018}, are equally capable of explaining the data. In the right column of Figure \ref{FullLightCurve}, we show the Lomb-Scargle periodograms, for the real and simulated observations, during the same time period. The  synthetic observations that we used as input for the periodograms were sampled at the cadence of the \cite{bel18} photometry. We show the periodogram of a flat photometric light curve sampled at the same cadence, to demonstrate that the long-period features are  artifacts of the window function.

For each model, we also computed light curves that had constant albedo, varying surface colors, and varying surface albedo. We found that the asymmetric surface variations with depths of order $\sim.5a$ and widths $\sim.4a$ could account for up to a $25\%$  increase in the magnitude of the oscillations in the light curve, and can explain the finer-scale variations in the observed photometric light curve. These  variations  had no distinguishable effect on the resultant periodograms.

Our model shows the general overall consistency of several example axis ratios with the data (notably 9:4:1), but we have not carried out an optimization for a precise shape and rotation model. Given (i) the relatively small photometric data set, (ii) the likely presence of chaotic tumbling, (iii) unknown deviations from ellipsoidal geometry, and (iv) potential albedo variations on the object’s surface, there is little indication of need for change in the current literature consensus \citep[e.g.][]{Mcneill2018} that the aspect ratio is high, likely $\gtrsim 5$:$1$, and that the motion may involve non-principal axis rotation \citep[e.g.][]{bel18}.

\section{Discussion}

`Oumuamua's $a\sim 5A_{\rm ng}P^2/4\pi^2 \sim 260\,{\rm m}$ long-axis size implied by its light-curve period and its acceleration is fully independent of the long-axis dimension inferred from its brightness.  If this agreement is not accidental, it is a point of evidence in favor of \citet{mic18}'s outgassing comet model, which in turn supports the hypothesis that `Oumuamua was ejected from the outer regions of a protoplanetary disk by a protoplanet having $v_{\rm esc}/v_{\rm orb}\sim (M_{\rm pl}a_{\rm orb}/M_{\star}R_{\rm pl})^{1/2}>1$.

`Oumuamua's acceleration in this scenario requires a highly volatile composition. Assuming that the acceleration-producing jet vented at $v_{\rm jet}\sim 3\times10^4\,{\rm cm\,s^{-1}}$, to order of magnitude, the overall mass loss rate was of order $\dot{m}=M A_{\rm ng}/v_{\rm jet}\sim 10^4\,{\rm g\,s^{-1}}$, where the mass, $M$ of `Oumuamua is of order $M\sim10^{12}\,{\rm g}$.  Assuming a pure water vapor jet, the required outflow rate is of order $Q[{\rm H_2O}]=3\times10^{26}\, {\rm s^{-1}}$. No direct measurements of water out-gassing were made during `Oumuamua's passage. \citet{Park2018} report upper limits on the ${\rm H_2O}$ dissociation product $Q[{\rm OH}]<1.7\times10^{27}\, {\rm \,s^{-1}}$, which is not in conflict with the required water production rate. Substantially tighter limits do, however, exist on the outflux of carbon-containing gasses $Q[{\rm CO_2}]<9\times10^{22}\, {\rm s^{-1}}$ \citep{Trilling2018}, $Q[{\rm CO}]<9\times10^{23}\, {\rm s^{-1}}$ \citep{Trilling2018}, and $Q[{\rm CN}]<2\times10^{22}\, {\rm s^{-1}}$ \citep{Ye2017}. If our elaboration of the jet model is correct, the combined detection limits imply a low C/O ratio for `Oumuamua's volatile component.  Assuming the upper limits quoted above as compositions, the nominal C/O ratio is only $0.003$. Clearly, such material is not pristine, with unknown physical processing required to produce the purification of the water ice. We note that some solar system comets are known to display very low abundances of carbon-containing molecules. For example, Comet 96P/Macholz 1 showed a ratio $[{\rm CN}/{\rm OH}]$ 72$\times$ smaller than average, and also showed $[{\rm C_2}/{\rm OH}]$ and $[{\rm C_3}/{\rm OH}]$ ratios $8\times$ and $19\times$ smaller than average respectively \citep{Schleicher2008}.

Sustained for the $\tau\sim100\,{\rm d}$ duration of its trajectory through the inner solar system,  Oumuamua's acceleration implies a total mass loss of $m=\dot{m} \tau \sim 10^{11}\,{\rm g}$, or $\sim 10\%$ of the total mass. `Oumuamua has thus likely not spent much time in close proximity to any star since its formation.
 
If `Oumuamua is indeed an ejected protoplanetary disk or exo-Oort cloud object \citep{Jewitt2017},  it implies a size-frequency distribution skewed more toward smaller bodies than expected from studies of the Solar System \citep[see, e.g.][]{Moro2009}. Both of these formation scenarios suggest a high-occurrence fraction for long-period sub-Jovian planets \citep{Laughlin2017}; either by a direct ejection of a planetesimal via a scattering event, or an ejection into an exo-Oort cloud and a subsequent stellar encounter or post-main sequence mass loss. Such a planet population is consistent with the one inferred from the gap structures that ALMA commonly observes in protostellar disks \citep{Andrews2018,Zhang2018},  whose ubiquity suggests that 50\% or more of young stars may contain $M\sim M_{\rm Nep}$ planets at large stellocentric distances.

We close by emphasizing that we have not fully solved the 'Oumuamua puzzle. In addition to the gas composition problem discussed above,
\citet{moro2018} and \citet{moro2019a} found that only very steep power law size distributions can achieve agreement with the number density of `Oumuamua-like objects derived by \citet{Do2018}, based on the aggregate Pan-STARRS search volume. The recent detection of a 1.2km KBO via occultation \citep{Arimatsu2019} suggests that very small bodies may be more numerous in the Kuiper Belt than often assumed, but like `Oumuamua, it presents a population of one. 

Only further work, in the form of further occultation surveys, the survey efforts of LSST, and eventually, \textit{in situ} sampling missions, will begin to unravel the mysteries that `Oumuamua's passage has provoked.

\section{acknowledgments}

This work was supported by the NASA Astrobiology Institute under Cooperative Agreement Notice NNH13ZDA017C issued through the Science Mission Directorate. We acknowledge support from the NASA Astrobiology Institute through a cooperative agreement between NASA Ames Research Center and Yale University.  We thank Hanno Rein and Karen Meech for useful conversations, and we thank Avi Loeb for finding a numerical error in an earlier draft of this manuscript.

\bibliographystyle{apj}
\bibliography{refs}

\begin{thebibliography}{}
\expandafter\ifx\csname natexlab\endcsname\relax\def\natexlab#1{#1}\fi

\bibitem[{{Andrews} {et~al.}(2018){Andrews}, {Huang}, {P{\'e}rez}, {Isella},
  {Dullemond}, {Kurtovic}, {Guzm{\'a}n}, {Carpenter}, {Wilner}, {Zhang}, {Zhu},
  {Birnstiel}, {Bai}, {Benisty}, {Hughes}, {{\"O}berg}, \&
  {Ricci}}]{Andrews2018}
{Andrews}, S.~M., {Huang}, J., {P{\'e}rez}, L.~M., {et~al.} 2018, \apjl, 869,
  L41

\bibitem[{{Arimatsu} {et~al.}(2019){Arimatsu}, {Tsumura}, {Usui}, {Shinnaka},
  {Ichikawa}, {Ootsubo}, {Kotani}, {Wada}, {Nagase}, \&
  {Watanabe}}]{Arimatsu2019}
{Arimatsu}, K., {Tsumura}, K., {Usui}, F., {et~al.} 2019, Nature Astronomy,
  doi:10.1038/s41550-018-0685-8

\bibitem[{{Belton} {et~al.}(2018){Belton}, {Hainaut}, {Meech}, {Mueller},
  {Kleyna}, {Weaver}, {Buie}, {Drahus}, {Guzik}, {Wainscoat}, {Waniak},
  {Handzlik}, {Kurowski}, {Xu}, {Sheppard}, {Micheli}, {Ebeling}, \&
  {Keane}}]{bel18}
{Belton}, M.~J.~S., {Hainaut}, O.~R., {Meech}, K.~J., {et~al.} 2018, \apjl,
  856, L21

\bibitem[{{Bialy} \& {Loeb}(2018)}]{Bialy2018}
{Bialy}, S., \& {Loeb}, A. 2018, ArXiv e-prints, arXiv:1810.11490

\bibitem[{{Cook} {et~al.}(2016){Cook}, {Ragozzine}, {Granvik}, \&
  {Stephens}}]{Cook2016}
{Cook}, N.~V., {Ragozzine}, D., {Granvik}, M., \& {Stephens}, D.~C. 2016, \apj,
  825, 51

\bibitem[{{Do} {et~al.}(2018){Do}, {Tucker}, \& {Tonry}}]{Do2018}
{Do}, A., {Tucker}, M.~A., \& {Tonry}, J. 2018, \apjl, 855, L10

\bibitem[{{Drahus} {et~al.}(2018){Drahus}, {Guzik}, {Waniak}, {Handzlik},
  {Kurowski}, \& {Xu}}]{dra18}
{Drahus}, M., {Guzik}, P., {Waniak}, W., {et~al.} 2018, Nature Astronomy, 2,
  407

\bibitem[{{Fraser} {et~al.}(2018){Fraser}, {Pravec}, {Fitzsimmons}, {Lacerda},
  {Bannister}, {Snodgrass}, \& {Smoli{\'c}}}]{fra18}
{Fraser}, W.~C., {Pravec}, P., {Fitzsimmons}, A., {et~al.} 2018, Nature
  Astronomy, 2, 383

\bibitem[{{Goldstein}(1950)}]{gol50}
{Goldstein}, H. 1950, {Classical mechanics}

\bibitem[{{Jewitt} {et~al.}(2017){Jewitt}, {Luu}, {Rajagopal}, {Kotulla},
  {Ridgway}, {Liu}, \& {Augusteijn}}]{Jewitt2017}
{Jewitt}, D., {Luu}, J., {Rajagopal}, J., {et~al.} 2017, \apjl, 850, L36

\bibitem[{{Kramer} \& {Laeuter}(2019)}]{kra19}
{Kramer}, T., \& {Laeuter}, M. 2019, arXiv e-prints, arXiv:1902.02701

\bibitem[{{Laughlin} \& {Batygin}(2017)}]{Laughlin2017}
{Laughlin}, G., \& {Batygin}, K. 2017, Research Notes of the American
  Astronomical Society, 1, 43

\bibitem[{{Lee} {et~al.}(2015){Lee}, {von Allmen}, {Allen}, {Beaudin}, {Biver},
  {Bockel{\'e}e-Morvan}, {Choukroun}, {Crovisier}, {Encrenaz}, {Frerking},
  {Gulkis}, {Hartogh}, {Hofstadter}, {Ip}, {Janssen}, {Jarchow}, {Keihm},
  {Lellouch}, {Leyrat}, {Rezac}, {Schloerb}, {Spilker}, {Gaskell}, {Jorda},
  {Keller}, \& {Sierks}}]{Lee2015}
{Lee}, S., {von Allmen}, P., {Allen}, M., {et~al.} 2015, \aap, 583, A5

\bibitem[{{Mamajek}(2017)}]{mam17}
{Mamajek}, E. 2017, Research Notes of the American Astronomical Society, 1, 21

\bibitem[{{Marsden} {et~al.}(1973){Marsden}, {Sekanina}, \& {Yeomans}}]{mar73}
{Marsden}, B.~G., {Sekanina}, Z., \& {Yeomans}, D.~K. 1973, \aj, 78, 211

\bibitem[{{McNeill} {et~al.}(2018){McNeill}, {Trilling}, \&
  {Mommert}}]{Mcneill2018}
{McNeill}, A., {Trilling}, D.~E., \& {Mommert}, M. 2018, \apjl, 857, L1

\bibitem[{{Meech} {et~al.}(2017){Meech}, {Weryk}, {Micheli}, {Kleyna},
  {Hainaut}, {Jedicke}, {Wainscoat}, {Chambers}, {Keane}, {Petric}, {Denneau},
  {Magnier}, {Berger}, {Huber}, {Flewelling}, {Waters}, {Schunova-Lilly}, \&
  {Chastel}}]{mee17}
{Meech}, K.~J., {Weryk}, R., {Micheli}, M., {et~al.} 2017, \nat, 552, 378

\bibitem[{{Micheli} {et~al.}(2018){Micheli}, {Farnocchia}, {Meech}, {Buie},
  {Hainaut}, {Prialnik}, {Sch{\"o}rghofer}, {Weaver}, {Chodas}, {Kleyna},
  {Weryk}, {Wainscoat}, {Ebeling}, {Keane}, {Chambers}, {Koschny}, \&
  {Petropoulos}}]{mic18}
{Micheli}, M., {Farnocchia}, D., {Meech}, K.~J., {et~al.} 2018, \nat, 559, 223

\bibitem[{{Moore} {et~al.}(1983){Moore}, {Donn}, {Khanna}, \&
  {A'Hearn}}]{Moore1983}
{Moore}, M.~H., {Donn}, B., {Khanna}, R., \& {A'Hearn}, M.~F. 1983, \icarus,
  54, 388

\bibitem[{{Moro-Mart{\'{\i}}n}(2018)}]{moro2018}
{Moro-Mart{\'{\i}}n}, A. 2018, \apj, 866, 131

\bibitem[{{Moro-Mart{\'{\i}}n}(2019{\natexlab{a}})}]{moro2019b}
---. 2019{\natexlab{a}}, arXiv e-prints, arXiv:1902.04100

\bibitem[{{Moro-Mart{\'{\i}}n}(2019{\natexlab{b}})}]{moro2019a}
---. 2019{\natexlab{b}}, \aj, 157, 86

\bibitem[{{Moro-Mart{\'{\i}}n} {et~al.}(2009){Moro-Mart{\'{\i}}n}, {Turner}, \&
  {Loeb}}]{Moro2009}
{Moro-Mart{\'{\i}}n}, A., {Turner}, E.~L., \& {Loeb}, A. 2009, \apj, 704, 733

\bibitem[{{Park} {et~al.}(2018){Park}, {Pisano}, {Lazio}, {Chodas}, \&
  {Naidu}}]{Park2018}
{Park}, R.~S., {Pisano}, D.~J., {Lazio}, T.~J.~W., {Chodas}, P.~W., \& {Naidu},
  S.~P. 2018, \aj, 155, 185

\bibitem[{{Rafikov}(2018)}]{Rafikov2018}
{Rafikov}, R.~R. 2018, \apjl, 867, L17

\bibitem[{{Rein}(2010)}]{Rein2010}
{Rein}, H. 2010, PhD thesis, PhD Thesis, 2010

\bibitem[{Rohatgi(2017)}]{ank17}
Rohatgi, A. 2017, {WebPlotDigitizer}, https://automeris.io/WebPlotDigitizer

\bibitem[{{Schleicher}(2008)}]{Schleicher2008}
{Schleicher}, D.~G. 2008, \aj, 136, 2204

\bibitem[{{Trilling} {et~al.}(2018){Trilling}, {Mommert}, {Hora}, {Farnocchia},
  {Chodas}, {Giorgini}, {Smith}, {Carey}, {Lisse}, {Werner}, {McNeill},
  {Chesley}, {Emery}, {Fazio}, {Fernandez}, {Harris}, {Marengo}, {Mueller},
  {Roegge}, {Smith}, {Weaver}, {Meech}, \& {Micheli}}]{Trilling2018}
{Trilling}, D.~E., {Mommert}, M., {Hora}, J.~L., {et~al.} 2018, \aj, 156, 261

\bibitem[{Williams(2017)}]{williams2017}
Williams, G.~V. 2017, MPEC 2017-U181: COMET C/2017 U1 (PANSTARRS)

\bibitem[{{Ye} {et~al.}(2017){Ye}, {Zhang}, {Kelley}, \& {Brown}}]{Ye2017}
{Ye}, Q.-Z., {Zhang}, Q., {Kelley}, M.~S.~P., \& {Brown}, P.~G. 2017, \apjl,
  851, L5

\bibitem[{{Zhang} {et~al.}(2018){Zhang}, {Zhu}, {Huang}, {Guzm{\'a}n},
  {Andrews}, {Birnstiel}, {Dullemond}, {Carpenter}, {Isella}, {P{\'e}rez},
  {Benisty}, {Wilner}, {Baruteau}, {Bai}, \& {Ricci}}]{Zhang2018}
{Zhang}, S., {Zhu}, Z., {Huang}, J., {et~al.} 2018, \apjl, 869, L47

\end{thebibliography}


\begin{thebibliography}{}


\end{thebibliography}
\end{document}